\def\sphinxdocclass{IEEEtran}
\documentclass[letterpaper,10pt,english,conference,onecolumn]{sphinxhowto}
\ifdefined\pdfpxdimen
   \let\sphinxpxdimen\pdfpxdimen\else\newdimen\sphinxpxdimen
\fi \sphinxpxdimen=.75bp\relax
\ifdefined\pdfimageresolution
    \pdfimageresolution= \numexpr \dimexpr1in\relax/\sphinxpxdimen\relax
\fi
\PassOptionsToPackage{bookmarksdepth=5}{hyperref}

\PassOptionsToPackage{booktabs}{sphinx}
\PassOptionsToPackage{colorrows}{sphinx}

\PassOptionsToPackage{warn}{textcomp}
\usepackage[utf8]{inputenc}
\ifdefined\DeclareUnicodeCharacter
  \ifdefined\DeclareUnicodeCharacterAsOptional
    \def\sphinxDUC#1{\DeclareUnicodeCharacter{"#1}}
  \else
    \let\sphinxDUC\DeclareUnicodeCharacter
  \fi
  \sphinxDUC{00A0}{\nobreakspace}
  \sphinxDUC{2500}{\sphinxunichar{2500}}
  \sphinxDUC{2502}{\sphinxunichar{2502}}
  \sphinxDUC{2514}{\sphinxunichar{2514}}
  \sphinxDUC{251C}{\sphinxunichar{251C}}
  \sphinxDUC{2572}{\textbackslash}
\fi
\usepackage{cmap}
\usepackage[T1]{fontenc}
\usepackage{amsmath,amssymb,amstext}
\usepackage{babel}

\usepackage{tgtermes}
\usepackage{tgheros}

\usepackage[Bjarne]{fncychap}
\usepackage[,nonumfigreset,mathnumfig]{sphinx}

\fvset{fontsize=\small}
\usepackage{geometry}
\usepackage{qrcode,graphicx,calc,amsthm,etoolbox,flushend}
\usepackage{fontawesome}

\usepackage{hyperref}
\usepackage{hypcap}% it must be loaded after hyperref.
\addto\captionsenglish{}

\usepackage{sphinxmessages}

\patchcmd{\thebibliography}{\addcontentsline{toc}{section}{\refname}}{}{}{}
\newtheorem{example}{Example}
\def\subparagraph{} % because IEEE classes don't define this, but titlesec assumes it's present

\title{The MQT Handbook\\{\Large A Summary of Design Automation Tools and\\ Software for Quantum Computing}}
\date{May 16, 2024}
\release{1.0.0}
\author{Robert Wille,\\%\hspace{3mm}%
		Lucas Berent,
		Tobias Forster,
		Jagatheesan Kunasaikaran,
		Kevin Mato,
		Tom Peham,
		Nils Quetschlich,
		Damian Rovara,
		Aaron Sander,
		Ludwig Schmid,
		Daniel Schönberger,
		Yannick Stade,\\%\hspace{3mm}%
		Lukas Burgholzer\\%\hspace{3mm}%
		Chair for Design Automation, Technical University of Munich, Germany\\%
		\href{mailto:quantum.cda@xcit.tum.de}{quantum.cda@xcit.tum.de}}
\newcommand{\sphinxlogo}{\sphinxincludegraphics{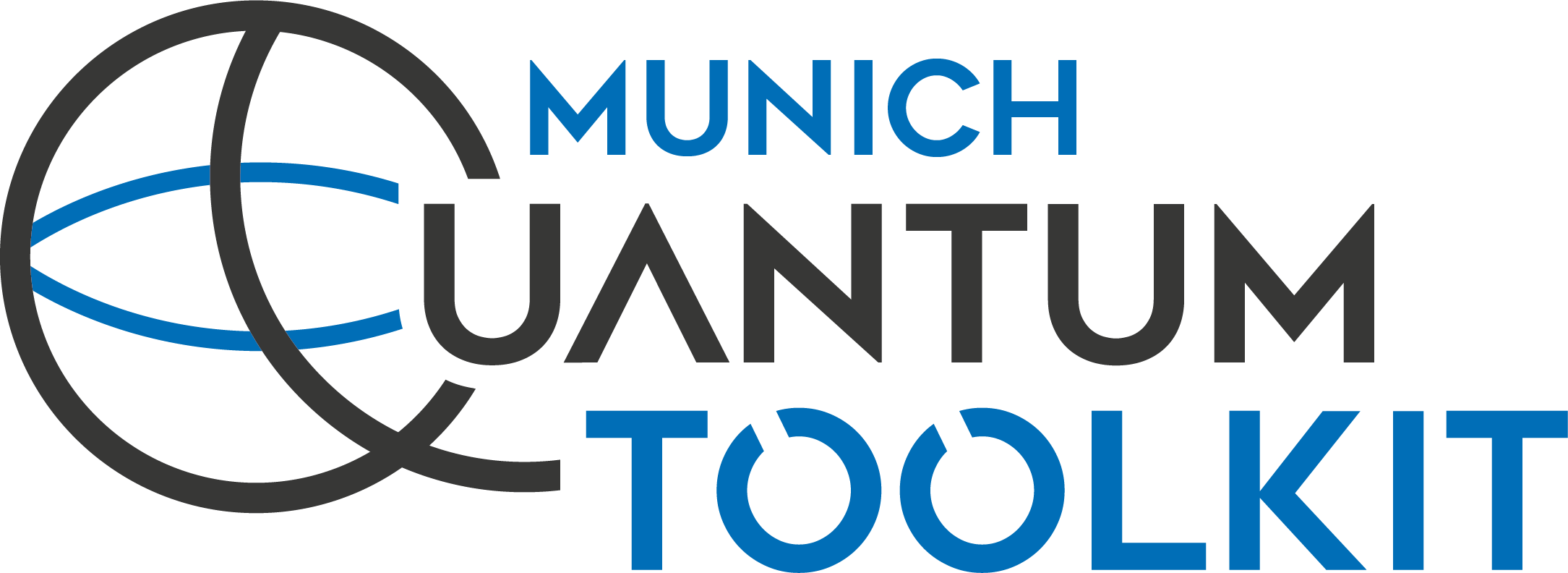}\par}
\renewcommand{\releasename}{Version}
\makeindex
\begin{document}

\ifdefined\shorthandoff
  \ifnum\catcode`\=\string=\active\shorthandoff{=}\fi
  \ifnum\catcode`\"=\active\shorthandoff{"}\fi
\fi

\pagestyle{empty}
\sphinxmaketitle
\pagestyle{plain}

\pagestyle{normal}
\phantomsection\label{\detokenize{index::doc}}

\begin{abstract}

\sphinxAtStartPar
Quantum computers are becoming a reality and numerous quantum computing applications with a near\sphinxhyphen{}term perspective (e.g., for finance, chemistry, machine learning, and optimization) and with a long\sphinxhyphen{}term perspective (e.g., for cryptography or unstructured search) are currently being investigated.
However, designing and realizing potential applications for these devices in a scalable fashion requires automated, efficient, and user\sphinxhyphen{}friendly software tools that cater to the needs of end users, engineers, and physicists at every level of the entire quantum software stack.
Many of the problems to be tackled in that regard are similar to design problems from the classical realm for which sophisticated design automation tools have been developed in the previous decades.

\sphinxAtStartPar
The \sphinxstyleemphasis{\sphinxhref{https://mqt.readthedocs.io}{Munich Quantum Toolkit (MQT)}} is a collection of software tools for quantum computing developed by the \sphinxhref{https://www.cda.cit.tum.de/}{Chair for Design Automation} at the \sphinxhref{https://www.tum.de/}{Technical University of Munich} which explicitly utilizes this design automation expertise.
Our overarching objective is to provide solutions for design tasks across the entire quantum software stack.
This entails high\sphinxhyphen{}level support for end users in realizing their \sphinxstyleemphasis{applications}, efficient methods for the \sphinxstyleemphasis{classical simulation}, \sphinxstyleemphasis{compilation}, and \sphinxstyleemphasis{verification} of quantum circuits, tools for \sphinxstyleemphasis{quantum error correction}, support for \sphinxstyleemphasis{physical design}, and more.
These methods are supported by corresponding \sphinxstyleemphasis{data structures} (such as decision diagrams) and \sphinxstyleemphasis{core methods} (such as SAT encodings/solvers).
All of the developed tools are available as open\sphinxhyphen{}source implementations and are hosted on \sphinxhref{https://github.com/cda-tum}{github.com/cda\sphinxhyphen{}tum}.

\begin{sphinxadmonition}{note}{Note:}
\sphinxAtStartPar
A live version of this document is available at \sphinxhref{https://mqt.readthedocs.io}{mqt.readthedocs.io}.
\end{sphinxadmonition}

\end{abstract}

\sphinxtableofcontents

\sphinxstepscope

\section{Introduction}
\label{\detokenize{handbook/01_intro:introduction}}\label{\detokenize{handbook/01_intro::doc}}
\sphinxAtStartPar
Quantum computing has the potential to revolutionize many fields in the 21st century, such as cryptography {[}\hyperlink{cite.handbook/references:id65}{1}{]}, finance {[}\hyperlink{cite.handbook/references:id66}{2}{]}, chemistry {[}\hyperlink{cite.handbook/references:id67}{3}{]}, machine learning {[}\hyperlink{cite.handbook/references:id68}{4}{]}, and optimization {[}\hyperlink{cite.handbook/references:id69}{5}{]}. Over the past decade, numerous quantum computers from multiple providers based on different qubit technologies have been made publicly available.
However, the best hardware is only as good as the software available to realize corresponding applications on it—a lesson learned from the past decades of research on designing and developing classical circuits and systems.
Thanks to the software tools and methods for \sphinxstyleemphasis{Electronic Design Automation (EDA)}, we can create classical systems with a staggering amount of transistors and complex functionalities that we often take for granted.
These methods allow designers to efficiently and automatically handle the intricacies of such systems and optimize their performance.
Compared to that, most existing software solutions for quantum computing leave the decades of research on design automation methods underutilized.

\sphinxAtStartPar
The \sphinxstyleemphasis{Munich Quantum Toolkit (MQT)}, which is developed by the \sphinxhref{https://www.cda.cit.tum.de/}{Chair for Design Automation} at the \sphinxhref{https://www.tum.de/}{Technical University of Munich}, aims to leverage this latent potential by providing a collection of state\sphinxhyphen{}of\sphinxhyphen{}the\sphinxhyphen{}art design automation methods and software tools for quantum computing.
Our overarching objective is to provide solutions for design tasks across the entire quantum software stack.
This entails high\sphinxhyphen{}level support for end users in realizing their \sphinxstyleemphasis{applications} {[}\hyperlink{cite.handbook/references:id33}{6}, \hyperlink{cite.handbook/references:id34}{7}, \hyperlink{cite.handbook/references:id35}{8}, \hyperlink{cite.handbook/references:id74}{9}, \hyperlink{cite.handbook/references:id75}{10}, \hyperlink{cite.handbook/references:id36}{11}{]}, efficient methods for the \sphinxstyleemphasis{classical simulation} {[}\hyperlink{cite.handbook/references:id7}{12}, \hyperlink{cite.handbook/references:id10}{13}, \hyperlink{cite.handbook/references:id11}{14}, \hyperlink{cite.handbook/references:id21}{15}, \hyperlink{cite.handbook/references:id12}{16}, \hyperlink{cite.handbook/references:id14}{17}, \hyperlink{cite.handbook/references:id13}{18}, \hyperlink{cite.handbook/references:id15}{19}, \hyperlink{cite.handbook/references:id17}{20}, \hyperlink{cite.handbook/references:id61}{21}, \hyperlink{cite.handbook/references:id57}{22}, \hyperlink{cite.handbook/references:id22}{23}, \hyperlink{cite.handbook/references:id18}{24}, \hyperlink{cite.handbook/references:id19}{25}, \hyperlink{cite.handbook/references:id20}{26}{]}, \sphinxstyleemphasis{compilation} {[}\hyperlink{cite.handbook/references:id36}{11}, \hyperlink{cite.handbook/references:id24}{27}, \hyperlink{cite.handbook/references:id25}{28}, \hyperlink{cite.handbook/references:id38}{29}, \hyperlink{cite.handbook/references:id26}{30}, \hyperlink{cite.handbook/references:id27}{31}, \hyperlink{cite.handbook/references:id28}{32}, \hyperlink{cite.handbook/references:id29}{33}, \hyperlink{cite.handbook/references:id30}{34}, \hyperlink{cite.handbook/references:id31}{35}, \hyperlink{cite.handbook/references:id37}{36}, \hyperlink{cite.handbook/references:id41}{37}, \hyperlink{cite.handbook/references:id39}{38}, \hyperlink{cite.handbook/references:id40}{39}, \hyperlink{cite.handbook/references:id43}{40}, \hyperlink{cite.handbook/references:id42}{41}, \hyperlink{cite.handbook/references:id60}{42}, \hyperlink{cite.handbook/references:id59}{43}, \hyperlink{cite.handbook/references:id58}{44}, \hyperlink{cite.handbook/references:id55}{45}, \hyperlink{cite.handbook/references:id16}{46}{]}, and \sphinxstyleemphasis{verification} {[}\hyperlink{cite.handbook/references:id46}{47}, \hyperlink{cite.handbook/references:id47}{48}, \hyperlink{cite.handbook/references:id48}{49}, \hyperlink{cite.handbook/references:id49}{50}, \hyperlink{cite.handbook/references:id50}{51}, \hyperlink{cite.handbook/references:id52}{52}, \hyperlink{cite.handbook/references:id53}{53}, \hyperlink{cite.handbook/references:id51}{54}, \hyperlink{cite.handbook/references:id54}{55}{]} of quantum circuits, tools for \sphinxstyleemphasis{quantum error correction} {[}\hyperlink{cite.handbook/references:id62}{56}, \hyperlink{cite.handbook/references:id63}{57}, \hyperlink{cite.handbook/references:id64}{58}, \hyperlink{cite.handbook/references:id76}{59}{]}, support for \sphinxstyleemphasis{physical design} {[}\hyperlink{cite.handbook/references:id56}{60}{]}, and more.
In all these tools, we try to utilize \sphinxstyleemphasis{data structures} (such as decision diagrams {[}\hyperlink{cite.handbook/references:id78}{61}, \hyperlink{cite.handbook/references:id8}{62}{]} or the ZX\sphinxhyphen{}calculus {[}\hyperlink{cite.handbook/references:id79}{63}, \hyperlink{cite.handbook/references:id80}{64}{]}) and \sphinxstyleemphasis{core methods} (such as reasoning engines {[}\hyperlink{cite.handbook/references:id77}{65}{]}) to facilitate the efficient handling of quantum computations.
The proposed solutions demonstrate how utilizing design automation expertise can lead to improved efficiency, scalability, and reliability.
In particular, they illustrate the immense benefits of leveraging expertise in classical circuit and system design rather than starting from scratch.
All tools developed as part of the MQT are made available as open\sphinxhyphen{}source packages on \sphinxhref{https://github.com/cda-tum/}{github.com/cda\sphinxhyphen{}tum}.

\sphinxAtStartPar
In the following, we briefly summarize some of the core methods and tools (covering classical simulation, compilation, and verification of quantum circuits as well as benchmarking).
We particularly focus on how to use the tools, but additionally provide references and links that offer detailed descriptions of the underlying methods as well as summaries of corresponding case studies and evaluations demonstrating the benefits.

\sphinxstepscope

\section{Classical Simulation of Quantum Circuits}
\label{\detokenize{handbook/02_simulation:classical-simulation-of-quantum-circuits}}\label{\detokenize{handbook/02_simulation::doc}}
\sphinxAtStartPar
Performing a quantum computation (commonly described as a quantum circuit) entails evolving an initial quantum state by applying a sequence of operations (also called gates) and measuring the resulting system.
Eventually, the goal should obviously be to do that on a real device.
However, there are several important reasons for simulating the corresponding computations on a classical machine, particularly in the early stages of the design:
As long as no suitable devices are available (e.g., in terms of scale, feasible computation depth, or accuracy), classical simulations of quantum circuits still allow one to explore and test quantum applications, even if only on a limited scale.
However, also with further progress in the capabilities of the hardware platforms, classical simulation will remain an essential part of the quantum computing design process, since it additionally allows access to \sphinxstyleemphasis{all} amplitudes of a resulting quantum state in contrast to a real device that only probabilistically returns measurement results.
Moreover, classical simulation provides means to study quantum error correction as well as a baseline to estimate the advantage of quantum computers over classical computers.

\sphinxAtStartPar
The classical simulation of quantum circuits is commonly conducted by performing consecutive matrix\sphinxhyphen{}vector multiplication, which many simulators realize by storing a dense representation of the complete state vector in memory and evolving it correspondingly (see, e.g., {[}\hyperlink{cite.handbook/references:id2}{66}, \hyperlink{cite.handbook/references:id3}{67}, \hyperlink{cite.handbook/references:id4}{68}, \hyperlink{cite.handbook/references:id5}{69}, \hyperlink{cite.handbook/references:id6}{70}{]}) or by relying on tensor network methods (see, e.g., {[}\hyperlink{cite.handbook/references:id70}{71}, \hyperlink{cite.handbook/references:id71}{72}, \hyperlink{cite.handbook/references:id72}{73}, \hyperlink{cite.handbook/references:id73}{74}{]}).
This approach quickly becomes intractable due to the exponential growth of the quantum state with respect to the number of qubits—quickly rendering such simulations infeasible even on supercomputer clusters.
Simulation methodologies based on decision diagrams {[}\hyperlink{cite.handbook/references:id7}{12}, \hyperlink{cite.handbook/references:id8}{62}, \hyperlink{cite.handbook/references:id9}{75}{]} are a promising complementary approach that frequently allows reducing the required memory by exploiting redundancies in the simulated quantum state.

\sphinxAtStartPar
The \sphinxstyleemphasis{MQT} offers the classical quantum circuit simulator \sphinxstyleemphasis{DDSIM} that can be used to perform various quantum circuit simulation tasks based on using decision diagrams as a data structure.
This includes strong and weak simulation {[}\hyperlink{cite.handbook/references:id7}{12}, \hyperlink{cite.handbook/references:id10}{13}, \hyperlink{cite.handbook/references:id11}{14}{]}, approximation techniques {[}\hyperlink{cite.handbook/references:id21}{15}, \hyperlink{cite.handbook/references:id12}{16}{]}, noise\sphinxhyphen{}aware simulation {[}\hyperlink{cite.handbook/references:id14}{17}, \hyperlink{cite.handbook/references:id13}{18}, \hyperlink{cite.handbook/references:id15}{19}{]}, hybrid Schrödinger\sphinxhyphen{}Feynman techniques {[}\hyperlink{cite.handbook/references:id17}{20}{]}, support for dynamic circuits, the computation of expectation values {[}\hyperlink{cite.handbook/references:id61}{21}{]}, the simulation of mixed\sphinxhyphen{}dimensional systems {[}\hyperlink{cite.handbook/references:id57}{22}{]}, and more {[}\hyperlink{cite.handbook/references:id22}{23}, \hyperlink{cite.handbook/references:id18}{24}, \hyperlink{cite.handbook/references:id19}{25}, \hyperlink{cite.handbook/references:id20}{26}{]}.

\begin{example}

\sphinxAtStartPar
Consider the following listing that describes the quantum circuit for generating a three\sphinxhyphen{}qubit GHZ state (also shown in \hyperref[\detokenize{handbook/02_simulation:fig-ghz-circuit}]{Fig.\@ \ref{\detokenize{handbook/02_simulation:fig-ghz-circuit}}}):

\begin{sphinxuseclass}{cell}
\begin{sphinxuseclass}{tag_remove-output}
\begin{sphinxuseclass}{cell_input}
\begin{sphinxVerbatim}[commandchars=\\\{\},numbers=left,firstnumber=1,stepnumber=1]
\PYG{k+kn}{from} \PYG{n+nn}{qiskit} \PYG{k+kn}{import} \PYG{n}{QuantumCircuit}

\PYG{n}{circ} \PYG{o}{=} \PYG{n}{QuantumCircuit}\PYG{p}{(}\PYG{l+m+mi}{3}\PYG{p}{)}
\PYG{n}{circ}\PYG{o}{.}\PYG{n}{h}\PYG{p}{(}\PYG{l+m+mi}{2}\PYG{p}{)}
\PYG{n}{circ}\PYG{o}{.}\PYG{n}{cx}\PYG{p}{(}\PYG{l+m+mi}{2}\PYG{p}{,} \PYG{l+m+mi}{1}\PYG{p}{)}
\PYG{n}{circ}\PYG{o}{.}\PYG{n}{cx}\PYG{p}{(}\PYG{l+m+mi}{1}\PYG{p}{,} \PYG{l+m+mi}{0}\PYG{p}{)}
\PYG{n}{circ}\PYG{o}{.}\PYG{n}{measure\PYGZus{}all}\PYG{p}{(}\PYG{p}{)}
\end{sphinxVerbatim}

\end{sphinxuseclass}
\end{sphinxuseclass}
\end{sphinxuseclass}
\begin{sphinxuseclass}{cell}
\begin{sphinxuseclass}{tag_remove-input}
\begin{sphinxuseclass}{cell_output}
\begin{figure}[htb]
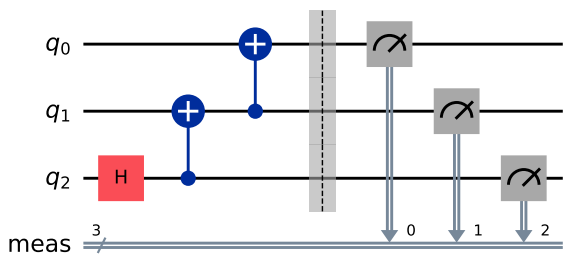

\centering
\capstart

\noindent\sphinxincludegraphics[width=0.500\linewidth]{{29b1b8d04fda7d8aa4ba632d99387148715247ca5d06b1a3a4286a34c8796dd0}.pdf}
\caption{Quantum circuit for generating a three\sphinxhyphen{}qubit GHZ state.}\label{\detokenize{handbook/02_simulation:fig-ghz-circuit}}\end{figure}

\end{sphinxuseclass}
\end{sphinxuseclass}
\end{sphinxuseclass}
\sphinxAtStartPar
This circuit can be classically simulated using DDSIM as a backend for IBM Qiskit:

\begin{sphinxuseclass}{cell}
\begin{sphinxuseclass}{cell_input}
\begin{sphinxVerbatim}[commandchars=\\\{\},numbers=left,firstnumber=1,stepnumber=1]
\PYG{k+kn}{from} \PYG{n+nn}{mqt}\PYG{n+nn}{.}\PYG{n+nn}{ddsim} \PYG{k+kn}{import} \PYG{n}{DDSIMProvider}

\PYG{n}{provider} \PYG{o}{=} \PYG{n}{DDSIMProvider}\PYG{p}{(}\PYG{p}{)}
\PYG{n}{backend} \PYG{o}{=} \PYG{n}{provider}\PYG{o}{.}\PYG{n}{get\PYGZus{}backend}\PYG{p}{(}\PYG{l+s+s2}{\PYGZdq{}}\PYG{l+s+s2}{qasm\PYGZus{}simulator}\PYG{l+s+s2}{\PYGZdq{}}\PYG{p}{)}
\PYG{n}{result} \PYG{o}{=} \PYG{n}{backend}\PYG{o}{.}\PYG{n}{run}\PYG{p}{(}\PYG{n}{circ}\PYG{p}{,} \PYG{n}{shots}\PYG{o}{=}\PYG{l+m+mi}{10000}\PYG{p}{)}\PYG{o}{.}\PYG{n}{result}\PYG{p}{(}\PYG{p}{)}
\PYG{n}{result}\PYG{o}{.}\PYG{n}{get\PYGZus{}counts}\PYG{p}{(}\PYG{p}{)}
\end{sphinxVerbatim}

\end{sphinxuseclass}
\begin{sphinxuseclass}{cell_output}
\begin{sphinxVerbatim}[commandchars=\\\{\}]
\PYGZob{}\PYGZsq{}000\PYGZsq{}: 4975, \PYGZsq{}111\PYGZsq{}: 5025\PYGZcb{}
\end{sphinxVerbatim}

\end{sphinxuseclass}
\end{sphinxuseclass}
\end{example}

\begin{minipage}[t]{0.76\linewidth}
\textbf{MQT DDSIM}\newline
\emph{Code:} \href{https://github.com/cda-tum/mqt-ddsim}{cda-tum/mqt-ddsim}\newline
\emph{Python Package:} \href{https://pypi.org/p/mqt.ddsim}{pypi.org/p/mqt.ddsim}\newline
\emph{Documentation:} \href{https://mqt.readthedocs.io/projects/ddsim}{mqt.rtfd.io/projects/ddsim}
\end{minipage}%
\hspace{1em}%
\begin{minipage}[t]{0.15\linewidth}
\raisebox{2mm -\dimexpr\depth}{%
\qrcode[height=1.75cm]{https://github.com/cda-tum/mqt-ddsim}
}
\end{minipage}

\sphinxstepscope

\section{Compilation of Quantum Circuits}
\label{\detokenize{handbook/03_compilation:compilation-of-quantum-circuits}}\label{\detokenize{handbook/03_compilation::doc}}
\sphinxAtStartPar
In today’s digital world, creating computer programs has become a crucial element of software development.
With the advent of high\sphinxhyphen{}level programming languages such as C++ or Python, the development process has become simpler and more efficient.
These languages enable developers to produce code that is more human\sphinxhyphen{}readable and understandable without having to worry about the underlying hardware’s low\sphinxhyphen{}level features.
But before these programs can be executed on a computer, they must be translated into machine code that the computer can process.
This procedure is known as \sphinxstyleemphasis{compilation}, and it entails converting high\sphinxhyphen{}level code into a binary format that the computer’s processor can directly execute.
By making it easier for more people to create computer programs, this has enabled the development of complex software applications that can run on many different platforms such as desktops, laptops, mobile phones or embedded devices.

\sphinxAtStartPar
Just as in classical computing, the design of quantum circuits and the development of quantum algorithms are fundamental in the development of quantum computing applications.
Quantum circuits are analogous to classical functions or programs in that they are a sequence of quantum gates that perform specific operations on quantum bits or qubits instead of classical bits.
Similarly to classical processors, quantum processors can only execute a certain set of native instructions, and they might further limit the qubits on which these operations might be applied.
Thus, any high\sphinxhyphen{}level quantum circuit (describing a quantum application) must be \sphinxstyleemphasis{compiled} into a representation that can be executed on the targeted device.
Most importantly, the resulting quantum circuit must only use gates that are native to the device on which it shall be executed.
If the device only has limited connectivity between its qubits, it must only apply gates to qubits that are connected on the device.
Naturally, the efficiency of this compilation process is critical because it can have a significant impact on the performance of the resulting quantum program.
Inefficient compilation can lead to longer execution times, higher error rates, and reduced accuracy in the final result.
Therefore, developing efficient compilation methods for quantum programs is essential to overcome the challenges of quantum computing and realize the potential of this technology.

\sphinxAtStartPar
In the following, we mainly focus on the \sphinxstyleemphasis{quantum circuit mapping} task.
This is a crucial step in the compilation flow, as it directly affects the feasibility and performance of the quantum circuit on a given device.
It involves finding a way to map the qubits of a quantum circuit to the qubits of a quantum device, while respecting the limited connectivity constraints of the device and minimizing the overhead of additional gates.
In most cases, it is not possible to statically define a mapping of the circuit’s qubits to the device’s qubits such that all gates of the circuit conform to the connectivity limitations of the device.
Consequently, this mapping has to change dynamically throughout the circuit.
This can be accomplished by using \sphinxstyleemphasis{SWAP} gates that allow the position of two logical qubits on the architecture to be interchanged.
However, since any additional gate increases the error rate and, hence, reduces the accuracy of the computation, it is vital to keep the number of additionally added gates as low as possible.
It has been shown that even this small part in the compilation flow is an NP\sphinxhyphen{}complete problem {[}\hyperlink{cite.handbook/references:id23}{76}{]}.

\sphinxAtStartPar
The \sphinxstyleemphasis{MQT} offers the quantum circuit mapping tool QMAP that allows one to generate circuits which satisfy all constraints given by the targeted architecture and, at the same time, keep the overhead in terms of additionally required quantum gates as low as possible.
More precisely, different approaches based on design automation techniques are provided, which are generic and can be easily configured for future architectures.
Among them is a heuristic, scalable solution for arbitrary circuits based on informed\sphinxhyphen{}search algorithms {[}\hyperlink{cite.handbook/references:id24}{27}, \hyperlink{cite.handbook/references:id25}{28}, \hyperlink{cite.handbook/references:id38}{29}{]} as well as a solution for obtaining mappings ensuring minimal overhead with respect to SWAP gate insertions {[}\hyperlink{cite.handbook/references:id26}{30}, \hyperlink{cite.handbook/references:id27}{31}{]}.

\sphinxAtStartPar
Additionally, the \sphinxstyleemphasis{MQT} offers many more methods for various compilation tasks, such as Clifford circuit synthesis {[}\hyperlink{cite.handbook/references:id28}{32}, \hyperlink{cite.handbook/references:id29}{33}{]}, determining optimal sub\sphinxhyphen{}architectures {[}\hyperlink{cite.handbook/references:id30}{34}{]}, compiler optimization {[}\hyperlink{cite.handbook/references:id36}{11}, \hyperlink{cite.handbook/references:id31}{35}, \hyperlink{cite.handbook/references:id37}{36}{]}, and compilation techniques for neutral atom technologies {[}\hyperlink{cite.handbook/references:id43}{40}, \hyperlink{cite.handbook/references:id42}{41}{]}, ion\sphinxhyphen{}trap shuttling {[}\hyperlink{cite.handbook/references:id41}{37}, \hyperlink{cite.handbook/references:id39}{38}, \hyperlink{cite.handbook/references:id40}{39}{]}, or multi\sphinxhyphen{}level quantum systems {[}\hyperlink{cite.handbook/references:id60}{42}, \hyperlink{cite.handbook/references:id59}{43}, \hyperlink{cite.handbook/references:id58}{44}, \hyperlink{cite.handbook/references:id55}{45}{]}.
Furthermore, it provides first automated methods regarding \sphinxstyleemphasis{Fault\sphinxhyphen{}Tolerant Quantum Computing} (FTQC) such as automatic circuit generation and evaluation for error\sphinxhyphen{}correcting codes {[}\hyperlink{cite.handbook/references:id16}{46}{]}.

\begin{example}

\sphinxAtStartPar
Assume we want to perform the computation from \hyperref[\detokenize{handbook/02_simulation:fig-ghz-circuit}]{Fig.\@ \ref{\detokenize{handbook/02_simulation:fig-ghz-circuit}}} on a five\sphinxhyphen{}qubit IBM quantum computer described by the coupling map shown in \hyperref[\detokenize{handbook/03_compilation:fig-device}]{Fig.\@ \ref{\detokenize{handbook/03_compilation:fig-device}}}.

\begin{sphinxuseclass}{cell}
\begin{sphinxuseclass}{tag_remove-input}
\begin{sphinxuseclass}{cell_output}
\begin{figure}[htb]
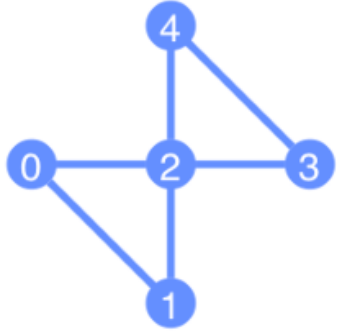

\centering
\capstart

\noindent\sphinxincludegraphics[width=0.300\linewidth]{{d99eb1940b96f07bf7a57f698e81ca6f4dd16ae030d014eab9aef625f15b56a4}.pdf}
\caption{Coupling map of a generic five\sphinxhyphen{}qubit IBM device.}\label{\detokenize{handbook/03_compilation:fig-device}}\end{figure}

\end{sphinxuseclass}
\end{sphinxuseclass}
\end{sphinxuseclass}
\sphinxAtStartPar
Then, mapping the circuit to that device merely requires the following lines of Python and results in the circuit shown in \hyperref[\detokenize{handbook/03_compilation:fig-ghz-circuit-mapped}]{Fig.\@ \ref{\detokenize{handbook/03_compilation:fig-ghz-circuit-mapped}}}.

\begin{sphinxuseclass}{cell}
\begin{sphinxuseclass}{cell_input}
\begin{sphinxVerbatim}[commandchars=\\\{\},numbers=left,firstnumber=1,stepnumber=1]
\PYG{k+kn}{from} \PYG{n+nn}{mqt}\PYG{n+nn}{.}\PYG{n+nn}{qmap} \PYG{k+kn}{import} \PYG{n+nb}{compile}
\PYG{k+kn}{from} \PYG{n+nn}{qiskit}\PYG{n+nn}{.}\PYG{n+nn}{providers}\PYG{n+nn}{.}\PYG{n+nn}{fake\PYGZus{}provider} \PYG{k+kn}{import} \PYG{n}{Fake5QV1}

\PYG{n}{backend} \PYG{o}{=} \PYG{n}{Fake5QV1}\PYG{p}{(}\PYG{p}{)}
\PYG{n}{circ\PYGZus{}mapped}\PYG{p}{,} \PYG{n}{results} \PYG{o}{=} \PYG{n+nb}{compile}\PYG{p}{(}\PYG{n}{circ}\PYG{p}{,} \PYG{n}{backend}\PYG{p}{)}
\end{sphinxVerbatim}

\end{sphinxuseclass}
\end{sphinxuseclass}
\begin{sphinxuseclass}{cell}
\begin{sphinxuseclass}{tag_remove-input}
\begin{sphinxuseclass}{cell_output}
\begin{figure}[htb]
\centering
\capstart

\noindent\sphinxincludegraphics[width=0.500\linewidth]{{b10d20620de6d4f6649e4a494316df4330748703ad0723f405d9c92858839834}.pdf}
\caption{Quantum circuit from \hyperref[\detokenize{handbook/02_simulation:fig-ghz-circuit}]{Fig.\@ \ref{\detokenize{handbook/02_simulation:fig-ghz-circuit}}} mapped to the five\sphinxhyphen{}qubit device shown in \hyperref[\detokenize{handbook/03_compilation:fig-device}]{Fig.\@ \ref{\detokenize{handbook/03_compilation:fig-device}}}.}\label{\detokenize{handbook/03_compilation:fig-ghz-circuit-mapped}}\end{figure}

\end{sphinxuseclass}
\end{sphinxuseclass}
\end{sphinxuseclass}
\end{example}

\begin{minipage}[t]{0.76\linewidth}
\textbf{MQT QMAP}\newline
\emph{Code:} \href{https://github.com/cda-tum/mqt-qmap}{cda-tum/mqt-qmap}\newline
\emph{Python Package:} \href{https://pypi.org/p/mqt.qmap}{pypi.org/p/mqt.qmap}\newline
\emph{Documentation:} \href{https://mqt.readthedocs.io/projects/qmap}{mqt.rtfd.io/projects/qmap}
\end{minipage}%
\hspace{1em}%
\begin{minipage}[t]{0.15\linewidth}
\raisebox{2mm -\dimexpr\depth}{%
\qrcode[height=1.75cm]{https://github.com/cda-tum/mqt-qmap}
}
\end{minipage}

\sphinxstepscope

\section{Verification of Quantum Circuits}
\label{\detokenize{handbook/04_verification:verification-of-quantum-circuits}}\label{\detokenize{handbook/04_verification::doc}}
\sphinxAtStartPar
Compiling quantum algorithms results in different representations of the considered functionality, which significantly differ in their basis operations and structure but are still supposed to be functionally equivalent.
As described in the previous section, even individual compilation tasks can be highly complex.
Consequently, checking whether the original functionality is indeed maintained throughout all these different abstractions becomes increasingly relevant in order to guarantee a consistent and error\sphinxhyphen{}free compilation flow.
This is similar to the classical realm, where descriptions at various levels of abstraction also exist.
These descriptions are verified using design automation expertise—resulting in efficient methods for verification to ensure the correctness of the design across different levels of abstraction {[}\hyperlink{cite.handbook/references:id44}{77}{]}.
However, since quantum circuits additionally employ quantum\sphinxhyphen{}physical effects such as superposition and entanglement, these methods cannot be used out of the box in the quantum realm.
Accordingly, verification of quantum circuits must be approached from a different perspective.
At first glance, these characteristics of quantum computing make verification much harder as for classical circuits and systems.
In fact, equivalence checking of quantum circuits has been proven to be a computationally hard problem {[}\hyperlink{cite.handbook/references:id45}{78}{]}.

\sphinxAtStartPar
At the same time, quantum circuits possess certain characteristics that offer remarkable potential for efficient equivalence checking that is not available in classical computing.
More precisely, consider two quantum circuits \(G=g_1,\dots,g_m\) and \(G'=g'_1,\dots,g'_n\) whose equivalence shall be checked.
Due to the inherent reversibility of quantum operations, the inverse of a quantum circuit can easily be computed by taking the complex conjugate of every gate and reversing the sequence of the gates in the circuit, i.e., \(G^{\prime -1}= (g'_n)^\dagger,\dots,(g'_1)^\dagger\).
If two circuits are equivalent, this allows for the conclusion that \(G\cdot G^{\prime -1} = I\), where \(I\) is the identity function.
Since the identity has the most compact representation for most data structures representing quantum functionality (e.g., linear with respect to the number of qubits in case of decision diagrams), the equivalence check can be simplified considerably.
Even complex circuits can be verified efficiently, if one manages to apply the gates of both circuits in a sequence that keeps the intermediate representation “close to the identity”.
Within the MQT, several methods and strategies were proposed that utilize this characteristic of quantum computations.
Eventually, this led to solutions that can verify the results of whole quantum compilation flows (such as IBM’s Qiskit) in negligible runtime—something we never managed for classical circuits and systems.

\sphinxAtStartPar
The \sphinxstyleemphasis{MQT} offers the quantum circuit equivalence checking tool QCEC which encompasses a comprehensive suite of efficient methods and automated tools for the verification of quantum circuits based on the ideas outlined in {[}\hyperlink{cite.handbook/references:id46}{47}, \hyperlink{cite.handbook/references:id47}{48}, \hyperlink{cite.handbook/references:id48}{49}, \hyperlink{cite.handbook/references:id49}{50}, \hyperlink{cite.handbook/references:id50}{51}, \hyperlink{cite.handbook/references:id52}{52}, \hyperlink{cite.handbook/references:id53}{53}, \hyperlink{cite.handbook/references:id51}{54}, \hyperlink{cite.handbook/references:id54}{55}{]}.
By this, an important step towards avoiding or substantially mitigating the emerge of a verification gap for quantum circuits is taken, i.e., a situation where the physical development of a technology substantially outperforms our ability to design suitable applications for it or to verify it.

\begin{example}

\sphinxAtStartPar
Verifying that the quantum circuit from \hyperref[\detokenize{handbook/03_compilation:fig-ghz-circuit-mapped}]{Fig.\@ \ref{\detokenize{handbook/03_compilation:fig-ghz-circuit-mapped}}} has been correctly compiled to the architecture from \hyperref[\detokenize{handbook/03_compilation:fig-device}]{Fig.\@ \ref{\detokenize{handbook/03_compilation:fig-device}}}, i.e., checking whether it still implements the functionality of the circuit shown in \hyperref[\detokenize{handbook/02_simulation:fig-ghz-circuit}]{Fig.\@ \ref{\detokenize{handbook/02_simulation:fig-ghz-circuit}}}, merely requires the following lines of Python:

\begin{sphinxuseclass}{cell}
\begin{sphinxuseclass}{cell_input}
\begin{sphinxVerbatim}[commandchars=\\\{\},numbers=left,firstnumber=1,stepnumber=1]
\PYG{k+kn}{from} \PYG{n+nn}{mqt}\PYG{n+nn}{.}\PYG{n+nn}{qcec} \PYG{k+kn}{import} \PYG{n}{verify}

\PYG{n}{result} \PYG{o}{=} \PYG{n}{verify}\PYG{p}{(}\PYG{n}{circ}\PYG{p}{,} \PYG{n}{circ\PYGZus{}mapped}\PYG{p}{)}
\PYG{n+nb}{print}\PYG{p}{(}\PYG{n}{result}\PYG{o}{.}\PYG{n}{equivalence}\PYG{p}{)}
\end{sphinxVerbatim}

\end{sphinxuseclass}
\begin{sphinxuseclass}{cell_output}
\begin{sphinxVerbatim}[commandchars=\\\{\}]
equivalent
\end{sphinxVerbatim}

\end{sphinxuseclass}
\end{sphinxuseclass}
\end{example}

\begin{minipage}[t]{0.76\linewidth}
\textbf{MQT QCEC}\newline
\emph{Code:} \href{https://github.com/cda-tum/mqt-qcec}{cda-tum/mqt-qcec}\newline
\emph{Python Package:} \href{https://pypi.org/p/mqt.qcec}{pypi.org/p/mqt.qcec}\newline
\emph{Documentation:} \href{https://mqt.readthedocs.io/projects/qcec}{mqt.rtfd.io/projects/qcec}
\end{minipage}%
\hspace{1em}%
\begin{minipage}[t]{0.15\linewidth}
\raisebox{2mm -\dimexpr\depth}{%
\qrcode[height=1.75cm]{https://github.com/cda-tum/mqt-qcec}
}
\end{minipage}

\sphinxstepscope

\section{Benchmarking Software and Design Automation Tools for Quantum Computing}
\label{\detokenize{handbook/05_benchmarking:benchmarking-software-and-design-automation-tools-for-quantum-computing}}\label{\detokenize{handbook/05_benchmarking::doc}}
\sphinxAtStartPar
Tools like the ones proposed above are key in order to support end users in the realization of their quantum computing applications.
And, thankfully, a huge variety of tools has been proposed in the past—with many more to come.
However, whenever such a quantum software tool is proposed, it is important to empirically evaluate its performance and to compare it to the state of the art.
For that purpose, proper benchmarks are needed.
To provide those, MQT Bench is proposed {[}\hyperlink{cite.handbook/references:id32}{79}{]}, which offers over \(70,000\) benchmarks on various abstraction levels (depending on what level the to\sphinxhyphen{}be\sphinxhyphen{}evaluated software tool operates on).
Having all those benchmarks in a single repository enables an increased comparability, reproducibility, and transparency.
To make the benchmarks as accessible as possible, MQT Bench comes as an easy\sphinxhyphen{}to\sphinxhyphen{}use website that is hosted at \sphinxhref{https://www.cda.cit.tum.de/mqtbench/}{www.cda.cit.tum.de/mqtbench/} and as a Python package available on \sphinxhref{https://pypi.org/project/mqt.bench/}{PyPI}.

\begin{example}

\sphinxAtStartPar
A larger version of the quantum circuit from \hyperref[\detokenize{handbook/02_simulation:fig-ghz-circuit}]{Fig.\@ \ref{\detokenize{handbook/02_simulation:fig-ghz-circuit}}} can easily be obtained programmatically from the MQT Bench Python package as follows:

\begin{sphinxuseclass}{cell}
\begin{sphinxuseclass}{cell_input}
\begin{sphinxVerbatim}[commandchars=\\\{\},numbers=left,firstnumber=1,stepnumber=1]
\PYG{k+kn}{from} \PYG{n+nn}{mqt}\PYG{n+nn}{.}\PYG{n+nn}{bench} \PYG{k+kn}{import} \PYG{n}{get\PYGZus{}benchmark}

\PYG{n}{circ} \PYG{o}{=} \PYG{n}{get\PYGZus{}benchmark}\PYG{p}{(}\PYG{l+s+s2}{\PYGZdq{}}\PYG{l+s+s2}{ghz}\PYG{l+s+s2}{\PYGZdq{}}\PYG{p}{,} \PYG{n}{circuit\PYGZus{}size}\PYG{o}{=}\PYG{l+m+mi}{8}\PYG{p}{,} \PYG{n}{level}\PYG{o}{=}\PYG{l+s+s2}{\PYGZdq{}}\PYG{l+s+s2}{alg}\PYG{l+s+s2}{\PYGZdq{}}\PYG{p}{)}
\end{sphinxVerbatim}

\end{sphinxuseclass}
\end{sphinxuseclass}
\begin{sphinxuseclass}{cell}
\begin{sphinxuseclass}{tag_remove-input}
\begin{sphinxuseclass}{cell_output}
\begin{figure}[htb]
\centering
\capstart

\noindent\sphinxincludegraphics[width=0.500\linewidth]{{1d98336be143faa6c4dbc6fbab7bdb8ec1dc6c7540d037ab445d28d1f76911f4}.pdf}
\caption{Larger version of the circuit from \hyperref[\detokenize{handbook/02_simulation:fig-ghz-circuit}]{Fig.\@ \ref{\detokenize{handbook/02_simulation:fig-ghz-circuit}}} obtained via MQT Bench.}\label{\detokenize{handbook/05_benchmarking:fig-ghz-circuit-bench}}\end{figure}

\end{sphinxuseclass}
\end{sphinxuseclass}
\end{sphinxuseclass}
\sphinxAtStartPar
This gives the circuit shown in \hyperref[\detokenize{handbook/05_benchmarking:fig-ghz-circuit-bench}]{Fig.\@ \ref{\detokenize{handbook/05_benchmarking:fig-ghz-circuit-bench}}}, which can then be used to evaluate the performance of a quantum software tool, e.g., to test how well the tool can simulate the circuit or how well it can compile it to a given architecture.

\end{example}

\begin{minipage}[t]{0.76\linewidth}
\textbf{MQT Bench}\newline
\emph{Code:} \href{https://github.com/cda-tum/mqt-bench}{cda-tum/mqt-bench}\newline
\emph{Python Package:} \href{https://pypi.org/p/mqt.bench}{pypi.org/p/mqt.bench}\newline
\emph{Documentation:} \href{https://mqt.readthedocs.io/projects/bench}{mqt.rtfd.io/projects/bench}
\end{minipage}%
\hspace{1em}%
\begin{minipage}[t]{0.15\linewidth}
\raisebox{2mm -\dimexpr\depth}{%
\qrcode[height=1.75cm]{https://github.com/cda-tum/mqt-bench}
}
\end{minipage}

\sphinxstepscope

\section{Open\sphinxhyphen{}Source Implementations}
\label{\detokenize{handbook/06_implementations:open-source-implementations}}\label{\detokenize{handbook/06_implementations::doc}}
\sphinxAtStartPar
All tools that have been developed as part of the \sphinxstyleemphasis{MQT} are publicly available on \sphinxhref{https://github.com/cda-tum\%7D\%7Bgithub.com/cda-tum}{github.com/cda\sphinxhyphen{}tum}.
Many of these tools are powered by MQT Core, which forms the backbone of the entire toolkit.
It features a comprehensive intermediate representation for quantum computations as well as a state\sphinxhyphen{}of\sphinxhyphen{}the\sphinxhyphen{}art decision diagram package for quantum computing and a dedicated ZX\sphinxhyphen{}calculus library.

\sphinxAtStartPar
All tools have been mainly implemented in C++, but strive to be as user\sphinxhyphen{}friendly as possible for the community.
Hence, push\sphinxhyphen{}button solutions are provided through Python bindings, pre\sphinxhyphen{}built Python wheels are available for all major platforms and Python versions, and all tools integrate natively with IBM’s Qiskit.
All tools are actively maintained and well documented.

\begin{minipage}[t]{0.76\linewidth}
\textbf{MQT Core}\newline
\emph{Code:} \href{https://github.com/cda-tum/mqt-core}{cda-tum/mqt-core}\newline
\emph{Python Package:} \href{https://pypi.org/p/mqt.core}{pypi.org/p/mqt.core}\newline
\emph{Documentation:} \href{https://mqt.readthedocs.io/projects/core}{mqt.rtfd.io/projects/core}
\end{minipage}%
\hspace{1em}%
\begin{minipage}[t]{0.15\linewidth}
\raisebox{2mm -\dimexpr\depth}{%
\qrcode[height=1.75cm]{https://github.com/cda-tum/mqt-core}
}
\end{minipage}

\sphinxstepscope

\section{Conclusions}
\label{\detokenize{handbook/07_conclusions:conclusions}}\label{\detokenize{handbook/07_conclusions::doc}}
\sphinxAtStartPar
Design automation tools and software have been crucial for the development of classical circuits and systems.
They enable faster and more reliable design cycles, reduce human errors, and allow for complex and large\sphinxhyphen{}scale designs.
In the domain of quantum computing, the corresponding design automation methods (which have been developed over the past decades) remain heavily underutilized.
The \sphinxstyleemphasis{Munich Quantum Toolkit (MQT)} makes substantial contributions towards leveraging this latent potential.
For many important design tasks, several methods and tools have been proposed that explicitly use design automation expertise while, at the same time, considering characteristics of quantum computing.
As the quantum computing landscape advances towards \sphinxstyleemphasis{Fault\sphinxhyphen{}Tolerant Quantum Computing} (FTQC), the \sphinxstyleemphasis{MQT} aims to support researchers, developers, and practitioners in the near\sphinxhyphen{}, middle\sphinxhyphen{}, and far\sphinxhyphen{}term future by providing a comprehensive suite of tools and methods.
\subsubsection*{Acknowledgments}

\sphinxAtStartPar
We thank everyone that contributed to the development of the \sphinxstyleemphasis{Munich Quantum Toolkit}.
Special thanks go to Alwin Zulehner, Stefan Hillmich, Thomas Grurl, Hartwig Bauer, Sarah Schneider, Smaran Adarsh, and Alexander Ploier for their specific contributions in the past.

\sphinxAtStartPar
The Munich Quantum Toolkit has been supported by the European Union’s Horizon 2020 research and innovation programme under the ERC Consolidator Grant (agreement No. 101001318), the NeQST Grant (agreement No. 101080086) and MILLENION (agreement No. 101114305).
It is part of the Munich Quantum Valley, which is supported by the Bavarian state government with funds from the Hightech Agenda Bayern Plus, and has been supported by the BMWK on the basis of a decision by the German Bundestag through project QuaST, as well as by the BMK, BMDW, the State of Upper Austria in the frame of the COMET program, and the QuantumReady project within Quantum Austria (managed by the FFG).

\sphinxstepscope

\begingroup
\renewcommand\section[1]{\endgroup}
\phantomsection

\section{References}
\label{\detokenize{handbook/references:references}}\label{\detokenize{handbook/references::doc}}
\begin{sphinxthebibliography}{10}
\bibitem[1]{handbook/references:id65}
\sphinxAtStartPar
P. W. Shor. Polynomial\sphinxhyphen{}time algorithms for prime factorization and discrete logarithms on a quantum computer. \sphinxstyleemphasis{SIAM Journal on Computing}, 1997. \sphinxhref{https://doi.org/10.1137/S0097539795293172}{doi:10.1137/S0097539795293172}.
\bibitem[2]{handbook/references:id66}
\sphinxAtStartPar
D. Egger, C. Gambella, J. Marecek, S. McFaddin, M. Mevissen, R. Raymond, A. Simonetto, S. Woerner, and E. Yndurain. Quantum Computing for Finance: State\sphinxhyphen{}of\sphinxhyphen{}the\sphinxhyphen{}Art and Future Prospects. \sphinxstyleemphasis{IEEE Transactions on Quantum Engineering}, 2020. \sphinxhref{https://doi.org/10.1109/TQE.2020.3030314}{doi:10.1109/TQE.2020.3030314}.
\bibitem[3]{handbook/references:id67}
\sphinxAtStartPar
S. McArdle, S. Endo, A. Aspuru\sphinxhyphen{}Guzik, S. C. Benjamin, and X. Yuan. Quantum computational chemistry. \sphinxstyleemphasis{Reviews of Modern Physics}, 92(1):015003, 2020. \sphinxhref{https://doi.org/10.1103/RevModPhys.92.015003}{doi:10.1103/RevModPhys.92.015003}.
\bibitem[4]{handbook/references:id68}
\sphinxAtStartPar
H.\sphinxhyphen{}Y. Huang, R. Kueng, G. Torlai, V. V. Albert, and J. Preskill. Provably efficient machine learning for quantum many\sphinxhyphen{}body problems. \sphinxstyleemphasis{Science}, 377(6613):eabk3333, 2022. \sphinxhref{https://doi.org/10.1126/science.abk3333}{doi:10.1126/science.abk3333}.
\bibitem[5]{handbook/references:id69}
\sphinxAtStartPar
S. Harwood, C. Gambella, D. Trenev, A. Simonetto, D. Bernal Neira, and D. Greenberg. Formulating and solving routing problems on quantum computers. \sphinxstyleemphasis{IEEE Transactions on Quantum Engineering}, 2:1\textendash{}17, 2021. \sphinxhref{https://doi.org/10.1109/TQE.2021.3049230}{doi:10.1109/TQE.2021.3049230}.
\bibitem[6]{handbook/references:id33}
\sphinxAtStartPar
N. Quetschlich, L. Burgholzer, and R. Wille. Towards an Automated Framework for Realizing Quantum Computing Solutions. In \sphinxstyleemphasis{Int\textquotesingle{}l Symp. on Multi\sphinxhyphen{}Valued Logic}. 2023. \sphinxhref{https://arxiv.org/abs/2210.14928}{arXiv:2210.14928}, \sphinxhref{https://doi.org/10.1109/ISMVL57333.2023.00035}{doi:10.1109/ISMVL57333.2023.00035}.
\bibitem[7]{handbook/references:id34}
\sphinxAtStartPar
N. Quetschlich, V. Koch, L. Burgholzer, and R. Wille. A hybrid classical quantum computing approach to the satellite mission planning problem. In \sphinxstyleemphasis{Int\textquotesingle{}l Conf. on Quantum Computing and Engineering}, volume 01, 642\textendash{}647. 2023. \sphinxhref{https://doi.org/10.1109/QCE57702.2023.00079}{doi:10.1109/QCE57702.2023.00079}.
\bibitem[8]{handbook/references:id35}
\sphinxAtStartPar
N. Quetschlich, L. Burgholzer, and R. Wille. Predicting Good Quantum Circuit Compilation Options. In \sphinxstyleemphasis{Int\textquotesingle{}l Conf. on Quantum Software}. 2023. \sphinxhref{https://arxiv.org/abs/2210.08027}{arXiv:2210.08027}, \sphinxhref{https://doi.org/10.1109/QSW59989.2023.00015}{doi:10.1109/QSW59989.2023.00015}.
\bibitem[9]{handbook/references:id74}
\sphinxAtStartPar
N. Quetschlich, M. Soeken, P. Murali, and R. Wille. Utilizing resource estimation for the development of quantum computing applications. 2024. \sphinxhref{https://arxiv.org/abs/2402.12434}{arXiv:2402.12434}.
\bibitem[10]{handbook/references:id75}
\sphinxAtStartPar
N. Quetschlich, F. J. Kiwit, M. A. Wolf, C. A. Riofrio, L. Burgholzer, A. Luckow, and R. Wille. Towards application\sphinxhyphen{}aware quantum circuit compilation. 2024. \sphinxhref{https://arxiv.org/abs/2404.12433}{arXiv:2404.12433}.
\bibitem[11]{handbook/references:id36}
\sphinxAtStartPar
N. Quetschlich, L. Burgholzer, and R. Wille. MQT Predictor: Automatic device selection with device\sphinxhyphen{}specific circuit compilation for quantum computing. 2023. \sphinxhref{https://arxiv.org/abs/2310.06889}{arXiv:2310.06889}.
\bibitem[12]{handbook/references:id7}
\sphinxAtStartPar
A. Zulehner and R. Wille. Advanced simulation of quantum computations. \sphinxstyleemphasis{IEEE Trans. on CAD of Integrated Circuits and Systems}, 2019. \sphinxhref{https://doi.org/10.1109/TCAD.2018.2834427}{doi:10.1109/TCAD.2018.2834427}.
\bibitem[13]{handbook/references:id10}
\sphinxAtStartPar
A. Zulehner and R. Wille. Matrix\sphinxhyphen{}Vector vs. Matrix\sphinxhyphen{}Matrix multiplication: Potential in DD\sphinxhyphen{}based simulation of quantum computations. In \sphinxstyleemphasis{Design, Automation and Test in Europe}. 2019. \sphinxhref{https://doi.org/10.23919/DATE.2019.8714836}{doi:10.23919/DATE.2019.8714836}.
\bibitem[14]{handbook/references:id11}
\sphinxAtStartPar
S. Hillmich, I. L. Markov, and R. Wille. Just like the real thing: Fast weak simulation of quantum computation. In \sphinxstyleemphasis{Design Automation Conf.} 2020. \sphinxhref{https://doi.org/10.1109/DAC18072.2020.9218555}{doi:10.1109/DAC18072.2020.9218555}.
\bibitem[15]{handbook/references:id21}
\sphinxAtStartPar
S. Hillmich, R. Kueng, I. L. Markov, and R. Wille. As accurate as needed, as efficient as possible: Approximations in DD\sphinxhyphen{}based quantum circuit simulation. In \sphinxstyleemphasis{Design, Automation and Test in Europe}. 2020. \sphinxhref{https://doi.org/10.23919/DATE51398.2021.9474034}{doi:10.23919/DATE51398.2021.9474034}.
\bibitem[16]{handbook/references:id12}
\sphinxAtStartPar
S. Hillmich, A. Zulehner, R. Kueng, I. L. Markov, and R. Wille. Approximating decision diagrams for quantum circuit simulation. \sphinxstyleemphasis{ACM Transactions on Quantum Computing}, 3(4):1\textendash{}21, 2022. \sphinxhref{https://doi.org/10.1145/3530776}{doi:10.1145/3530776}.
\bibitem[17]{handbook/references:id14}
\sphinxAtStartPar
T. Grurl, J. Fuß, and R. Wille. Considering decoherence errors in the simulation of quantum circuits using decision diagrams. In \sphinxstyleemphasis{Int\textquotesingle{}l Conf. on CAD}. 2020. \sphinxhref{https://doi.org/10.1145/3400302.3415622}{doi:10.1145/3400302.3415622}.
\bibitem[18]{handbook/references:id13}
\sphinxAtStartPar
T. Grurl, R. Kueng, J. Fuß, and R. Wille. Stochastic quantum circuit simulation using decision diagrams. In \sphinxstyleemphasis{Design, Automation and Test in Europe}. 2021. \sphinxhref{https://doi.org/10.23919/DATE51398.2021.9474135}{doi:10.23919/DATE51398.2021.9474135}.
\bibitem[19]{handbook/references:id15}
\sphinxAtStartPar
T. Grurl, J. Fuß, and R. Wille. Noise\sphinxhyphen{}aware quantum circuit simulation with decision diagrams. \sphinxstyleemphasis{IEEE Trans. on CAD of Integrated Circuits and Systems}, 42(3):860\textendash{}873, 2023. \sphinxhref{https://doi.org/10.1109/TCAD.2022.3182628}{doi:10.1109/TCAD.2022.3182628}.
\bibitem[20]{handbook/references:id17}
\sphinxAtStartPar
L. Burgholzer, H. Bauer, and R. Wille. Hybrid Schrödinger\sphinxhyphen{}Feynman simulation of quantum circuits with decision diagrams. In \sphinxstyleemphasis{Int\textquotesingle{}l Conf. on Quantum Computing and Engineering}. 2021. \sphinxhref{https://doi.org/10.1109/QCE52317.2021.00037}{doi:10.1109/QCE52317.2021.00037}.
\bibitem[21]{handbook/references:id61}
\sphinxAtStartPar
A. Sander, L. Burgholzer, and R. Wille. Towards hamiltonian simulation with decision diagrams. In \sphinxstyleemphasis{Int\textquotesingle{}l Conf. on Quantum Computing and Engineering}. 2023. \sphinxhref{https://arxiv.org/abs/2305.02337}{arXiv:2305.02337}, \sphinxhref{https://doi.org/10.1109/QCE57702.2023.00039}{doi:10.1109/QCE57702.2023.00039}.
\bibitem[22]{handbook/references:id57}
\sphinxAtStartPar
K. Mato, S. Hillmich, and R. Wille. Mixed\sphinxhyphen{}dimensional quantum circuit simulation with decision diagrams. In \sphinxstyleemphasis{Int\textquotesingle{}l Conf. on Quantum Computing and Engineering}. 2023. \sphinxhref{https://arxiv.org/abs/2308.12332}{arXiv:2308.12332}, \sphinxhref{https://doi.org/10.1109/QCE57702.2023.00112}{doi:10.1109/QCE57702.2023.00112}.
\bibitem[23]{handbook/references:id22}
\sphinxAtStartPar
S. Hillmich, A. Zulehner, and R. Wille. Concurrency in DD\sphinxhyphen{}based quantum circuit simulation. In \sphinxstyleemphasis{Asia and South Pacific Design Automation Conf.} 2020. \sphinxhref{https://doi.org/10.1109/ASP-DAC47756.2020.9045711}{doi:10.1109/ASP\sphinxhyphen{}DAC47756.2020.9045711}.
\bibitem[24]{handbook/references:id18}
\sphinxAtStartPar
L. Burgholzer, A. Ploier, and R. Wille. Exploiting arbitrary paths for the simulation of quantum circuits with decision diagrams. In \sphinxstyleemphasis{Design, Automation and Test in Europe}. 2022. \sphinxhref{https://doi.org/10.23919/DATE54114.2022.9774631}{doi:10.23919/DATE54114.2022.9774631}.
\bibitem[25]{handbook/references:id19}
\sphinxAtStartPar
L. Burgholzer, A. Ploier, and R. Wille. Simulation paths for quantum circuit simulation with decision diagrams: What to learn from tensor networks, and what not. \sphinxstyleemphasis{IEEE Trans. on CAD of Integrated Circuits and Systems}, 2022. \sphinxhref{https://arxiv.org/abs/2203.00703}{arXiv:2203.00703}, \sphinxhref{https://doi.org/10.1109/TCAD.2022.3197969}{doi:10.1109/TCAD.2022.3197969}.
\bibitem[26]{handbook/references:id20}
\sphinxAtStartPar
L. Burgholzer, R. Raymond, I. Sengupta, and R. Wille. Efficient construction of functional representations for quantum algorithms. In \sphinxstyleemphasis{Int\textquotesingle{}l Conf. of Reversible Computation}. 2021. \sphinxhref{https://doi.org/10.1007/978-3-030-79837-6\_14}{doi:10.1007/978\sphinxhyphen{}3\sphinxhyphen{}030\sphinxhyphen{}79837\sphinxhyphen{}6\_14}.
\bibitem[27]{handbook/references:id24}
\sphinxAtStartPar
A. Zulehner, A. Paler, and R. Wille. An efficient methodology for mapping quantum circuits to the IBM QX architectures. \sphinxstyleemphasis{IEEE Trans. on CAD of Integrated Circuits and Systems}, 2019. \sphinxhref{https://doi.org/10.1109/TCAD.2018.2846658}{doi:10.1109/TCAD.2018.2846658}.
\bibitem[28]{handbook/references:id25}
\sphinxAtStartPar
S. Hillmich, A. Zulehner, and R. Wille. Exploiting Quantum Teleportation in Quantum Circuit Mapping. In \sphinxstyleemphasis{Asia and South Pacific Design Automation Conf.}, 792\textendash{}797. 2021. \sphinxhref{https://doi.org/10.1145/3394885.3431604}{doi:10.1145/3394885.3431604}.
\bibitem[29]{handbook/references:id38}
\sphinxAtStartPar
A. Zulehner and R. Wille. Compiling SU(4) quantum circuits to IBM QX architectures. In \sphinxstyleemphasis{Asia and South Pacific Design Automation Conf.}, 185\textendash{}190. 2019. \sphinxhref{https://doi.org/10.1145/3287624.3287704}{doi:10.1145/3287624.3287704}.
\bibitem[30]{handbook/references:id26}
\sphinxAtStartPar
R. Wille, L. Burgholzer, and A. Zulehner. Mapping quantum circuits to IBM QX architectures using the minimal number of SWAP and H operations. In \sphinxstyleemphasis{Design Automation Conf.} 2019. \sphinxhref{https://doi.org/10.1145/3316781.3317859}{doi:10.1145/3316781.3317859}.
\bibitem[31]{handbook/references:id27}
\sphinxAtStartPar
L. Burgholzer, S. Schneider, and R. Wille. Limiting the search space in optimal quantum circuit mapping. In \sphinxstyleemphasis{Asia and South Pacific Design Automation Conf.} 2022. \sphinxhref{https://doi.org/10.1109/ASP-DAC52403.2022.9712555}{doi:10.1109/ASP\sphinxhyphen{}DAC52403.2022.9712555}.
\bibitem[32]{handbook/references:id28}
\sphinxAtStartPar
T. Peham, N. Brandl, R. Kueng, R. Wille, and L. Burgholzer. Depth\sphinxhyphen{}optimal synthesis of Clifford circuits with SAT solvers. In \sphinxstyleemphasis{Int\textquotesingle{}l Conf. on Quantum Computing and Engineering}. 2023. \sphinxhref{https://arxiv.org/abs/2305.01674}{arXiv:2305.01674}, \sphinxhref{https://doi.org/10.1109/QCE57702.2023.00095}{doi:10.1109/QCE57702.2023.00095}.
\bibitem[33]{handbook/references:id29}
\sphinxAtStartPar
S. Schneider, L. Burgholzer, and R. Wille. A SAT encoding for optimal Clifford circuit synthesis. In \sphinxstyleemphasis{Asia and South Pacific Design Automation Conf.} 2023. \sphinxhref{https://doi.org/10.1145/3566097.3567929}{doi:10.1145/3566097.3567929}.
\bibitem[34]{handbook/references:id30}
\sphinxAtStartPar
T. Peham, L. Burgholzer, and R. Wille. On Optimal Subarchitectures for Quantum Circuit Mapping. \sphinxstyleemphasis{ACM Transactions on Quantum Computing}, 2023. \sphinxhref{https://arxiv.org/abs/2210.09321}{arXiv:2210.09321}, \sphinxhref{https://doi.org/10.1145/3593594}{doi:10.1145/3593594}.
\bibitem[35]{handbook/references:id31}
\sphinxAtStartPar
N. Quetschlich, L. Burgholzer, and R. Wille. Compiler Optimization for Quantum Computing Using Reinforcement Learning. In \sphinxstyleemphasis{Design Automation Conf.} 2023. \sphinxhref{https://arxiv.org/abs/2212.04508}{arXiv:2212.04508}, \sphinxhref{https://doi.org/10.1109/DAC56929.2023.10248002}{doi:10.1109/DAC56929.2023.10248002}.
\bibitem[36]{handbook/references:id37}
\sphinxAtStartPar
N. Quetschlich, L. Burgholzer, and R. Wille. Reducing the compilation time of quantum circuits using pre\sphinxhyphen{}compilation on the gate level. In \sphinxstyleemphasis{Int\textquotesingle{}l Conf. on Quantum Computing and Engineering}. 2023. \sphinxhref{https://arxiv.org/abs/2305.04941}{arXiv:2305.04941}, \sphinxhref{https://doi.org/10.1109/QCE57702.2023.00091}{doi:10.1109/QCE57702.2023.00091}.
\bibitem[37]{handbook/references:id41}
\sphinxAtStartPar
D. Schoenberger, S. Hillmich, M. Brandl, and R. Wille. Using Boolean Satisfiability for Exact Shuttling in Trapped\sphinxhyphen{}Ion Quantum Computers. In \sphinxstyleemphasis{Asia and South Pacific Design Automation Conf.} 2024. \sphinxhref{https://doi.org/10.1109/ASP-DAC58780.2024.10473902}{doi:10.1109/ASP\sphinxhyphen{}DAC58780.2024.10473902}.
\bibitem[38]{handbook/references:id39}
\sphinxAtStartPar
D. Schoenberger, S. Hillmich, M. Brandl, and R. Wille. Towards Cycle\sphinxhyphen{}based Shuttling for Trapped\sphinxhyphen{}Ion Quantum Computers. In \sphinxstyleemphasis{Design, Automation and Test in Europe}. 2024.
\bibitem[39]{handbook/references:id40}
\sphinxAtStartPar
D. Schoenberger, S. Hillmich, M. Brandl, and R. Wille. Shuttling for Scalable Trapped\sphinxhyphen{}Ion Quantum Computers. 2024. \sphinxhref{https://arxiv.org/abs/2402.14065}{arXiv:2402.14065}.
\bibitem[40]{handbook/references:id43}
\sphinxAtStartPar
L. Schmid, S. Park, and R. Wille. Hybrid Circuit Mapping: Leveraging the Full Spectrum of Computational Capabilities of Neutral Atom Quantum Computers. In \sphinxstyleemphasis{Design Automation Conf.} 2024.
\bibitem[41]{handbook/references:id42}
\sphinxAtStartPar
L. Schmid, D. Locher, M. Rispler, S. Blatt, J. Zeiher, M. Müller, and R. Wille. Computational Capabilities and Compiler Development for Neutral Atom Quantum Processors \sphinxhyphen{} Connecting Tool Developers and Hardware Experts. \sphinxstyleemphasis{Quantum Science and Technology}, 2024. \sphinxhref{https://doi.org/10.1088/2058-9565/ad33ac}{doi:10.1088/2058\sphinxhyphen{}9565/ad33ac}.
\bibitem[42]{handbook/references:id60}
\sphinxAtStartPar
K. Mato, M. Ringbauer, S. Hillmich, and R. Wille. Adaptive compilation of multi\sphinxhyphen{}level quantum operations. In \sphinxstyleemphasis{Int\textquotesingle{}l Conf. on Quantum Computing and Engineering}, 484\textendash{}491. 2022. \sphinxhref{https://doi.org/10.1109/QCE53715.2022.00070}{doi:10.1109/QCE53715.2022.00070}.
\bibitem[43]{handbook/references:id59}
\sphinxAtStartPar
K. Mato, M. Ringbauer, S. Hillmich, and R. Wille. Compilation of entangling gates for high\sphinxhyphen{}dimensional quantum systems. In \sphinxstyleemphasis{Asia and South Pacific Design Automation Conf.}, 202\textendash{}208. 2023. \sphinxhref{https://doi.org/10.1145/3566097.3567930}{doi:10.1145/3566097.3567930}.
\bibitem[44]{handbook/references:id58}
\sphinxAtStartPar
K. Mato, S. Hillmich, and R. Wille. Compression of qubit circuits: Mapping to mixed\sphinxhyphen{}dimensional quantum systems. In \sphinxstyleemphasis{Int\textquotesingle{}l Conf. on Quantum Software}, 155\textendash{}161. 2023. \sphinxhref{https://doi.org/10.1109/QSW59989.2023.00027}{doi:10.1109/QSW59989.2023.00027}.
\bibitem[45]{handbook/references:id55}
\sphinxAtStartPar
K. Mato, S. Hillmich, and R. Wille. Mixed\sphinxhyphen{}dimensional qudit state preparation using edge\sphinxhyphen{}weighted decision diagrams. In \sphinxstyleemphasis{Design Automation Conf.} 2024.
\bibitem[46]{handbook/references:id16}
\sphinxAtStartPar
T. Grurl, C. Pichler, J. Fuß, and R. Wille. Automatic Implementation and Evaluation of Error\sphinxhyphen{}Correcting Codes for Quantum Computing: An Open\sphinxhyphen{}Source Framework for Quantum Error Correction. In \sphinxstyleemphasis{VLSI Design}, 301\textendash{}306. 2023. \sphinxhref{https://doi.org/10.1109/VLSID57277.2023.00068}{doi:10.1109/VLSID57277.2023.00068}.
\bibitem[47]{handbook/references:id46}
\sphinxAtStartPar
L. Burgholzer and R. Wille. Advanced equivalence checking for quantum circuits. \sphinxstyleemphasis{IEEE Trans. on CAD of Integrated Circuits and Systems}, 2021. \sphinxhref{https://doi.org/10.1109/TCAD.2020.3032630}{doi:10.1109/TCAD.2020.3032630}.
\bibitem[48]{handbook/references:id47}
\sphinxAtStartPar
L. Burgholzer and R. Wille. Improved DD\sphinxhyphen{}based equivalence checking of quantum circuits. In \sphinxstyleemphasis{Asia and South Pacific Design Automation Conf.} 2020. \sphinxhref{https://doi.org/10.1109/ASP-DAC47756.2020.9045153}{doi:10.1109/ASP\sphinxhyphen{}DAC47756.2020.9045153}.
\bibitem[49]{handbook/references:id48}
\sphinxAtStartPar
L. Burgholzer and R. Wille. The power of simulation for equivalence checking in quantum computing. In \sphinxstyleemphasis{Design Automation Conf.} 2020. \sphinxhref{https://doi.org/10.1109/DAC18072.2020.9218563}{doi:10.1109/DAC18072.2020.9218563}.
\bibitem[50]{handbook/references:id49}
\sphinxAtStartPar
L. Burgholzer, R. Kueng, and R. Wille. Random stimuli generation for the verification of quantum circuits. In \sphinxstyleemphasis{Asia and South Pacific Design Automation Conf.} 2021. \sphinxhref{https://doi.org/10.1145/3394885.3431590}{doi:10.1145/3394885.3431590}.
\bibitem[51]{handbook/references:id50}
\sphinxAtStartPar
L. Burgholzer, R. Raymond, and R. Wille. Verifying results of the IBM Qiskit quantum circuit compilation flow. In \sphinxstyleemphasis{Int\textquotesingle{}l Conf. on Quantum Computing and Engineering}. 2020. \sphinxhref{https://doi.org/10.1109/QCE49297.2020.00051}{doi:10.1109/QCE49297.2020.00051}.
\bibitem[52]{handbook/references:id52}
\sphinxAtStartPar
T. Peham, L. Burgholzer, and R. Wille. Equivalence checking of parameterized quantum circuits: Verifying the compilation of variational quantum algorithms. In \sphinxstyleemphasis{Asia and South Pacific Design Automation Conf.} 2023. \sphinxhref{https://doi.org/10.1145/3566097.3567932}{doi:10.1145/3566097.3567932}.
\bibitem[53]{handbook/references:id53}
\sphinxAtStartPar
T. Peham, L. Burgholzer, and R. Wille. Equivalence checking of quantum circuits with the ZX\sphinxhyphen{}Calculus. \sphinxstyleemphasis{IEEE Journal on Emerging and Selected Topics in Circuits and Systems}, 2022. \sphinxhref{https://doi.org/10.1109/JETCAS.2022.3202204}{doi:10.1109/JETCAS.2022.3202204}.
\bibitem[54]{handbook/references:id51}
\sphinxAtStartPar
T. Peham, L. Burgholzer, and R. Wille. Equivalence checking paradigms in quantum circuit design: A case study. In \sphinxstyleemphasis{Design Automation Conf.} 2022. \sphinxhref{https://doi.org/10.1145/3489517.3530480}{doi:10.1145/3489517.3530480}.
\bibitem[55]{handbook/references:id54}
\sphinxAtStartPar
R. Wille and L. Burgholzer. Verification of Quantum Circuits. In A. Chattopadhyay, editor, \sphinxstyleemphasis{Handbook of Computer Architecture}, pages 1\textendash{}28. Springer Nature Singapore, Singapore, 2022. \sphinxhref{https://doi.org/10.1007/978-981-15-6401-7\_43-1}{doi:10.1007/978\sphinxhyphen{}981\sphinxhyphen{}15\sphinxhyphen{}6401\sphinxhyphen{}7\_43\sphinxhyphen{}1}.
\bibitem[56]{handbook/references:id62}
\sphinxAtStartPar
L. Berent, L. Burgholzer, P.\sphinxhyphen{}J. H. S. Derks, J. Eisert, and R. Wille. Decoding quantum color codes with MaxSAT. 2023. \sphinxhref{https://arxiv.org/abs/2303.14237}{arXiv:2303.14237}.
\bibitem[57]{handbook/references:id63}
\sphinxAtStartPar
L. Berent, L. Burgholzer, and R. Wille. Software tools for decoding quantum low\sphinxhyphen{}density parity check codes. In \sphinxstyleemphasis{Asia and South Pacific Design Automation Conf.} 2023. \sphinxhref{https://doi.org/10.1145/3566097.3567934}{doi:10.1145/3566097.3567934}.
\bibitem[58]{handbook/references:id64}
\sphinxAtStartPar
A. Strikis and L. Berent. Quantum low\sphinxhyphen{}density parity\sphinxhyphen{}check codes for modular architectures. \sphinxstyleemphasis{PRX Quantum}, 4(2):020321, 2023. \sphinxhref{https://doi.org/10.1103/PRXQuantum.4.020321}{doi:10.1103/PRXQuantum.4.020321}.
\bibitem[59]{handbook/references:id76}
\sphinxAtStartPar
L. Berent, T. Hillmann, J. Eisert, R. Wille, and J. Roffe. Analog information decoding of bosonic quantum LDPC codes. 2023. \sphinxhref{https://arxiv.org/abs/2311.01328}{arXiv:2311.01328}.
\bibitem[60]{handbook/references:id56}
\sphinxAtStartPar
J. Kunasaikaran, K. Mato, and R. Wille. A framework for the design and realization of alternative superconducting quantum architectures. In \sphinxstyleemphasis{Int\textquotesingle{}l Symp. on Multi\sphinxhyphen{}Valued Logic}. 2024.
\bibitem[61]{handbook/references:id78}
\sphinxAtStartPar
R. Wille, S. Hillmich, and B. Lukas. Tools for quantum computing based on decision diagrams. \sphinxstyleemphasis{ACM Transactions on Quantum Computing}, 2022. \sphinxhref{https://doi.org/10.1145/3491246}{doi:10.1145/3491246}.
\bibitem[62]{handbook/references:id8}
\sphinxAtStartPar
R. Wille, S. Hillmich, and L. Burgholzer. Decision Diagrams for Quantum Computing. In \sphinxstyleemphasis{Design Automation of Quantum Computers}. 2023. \sphinxhref{https://doi.org/10.1007/978-3-031-15699-1\_1}{doi:10.1007/978\sphinxhyphen{}3\sphinxhyphen{}031\sphinxhyphen{}15699\sphinxhyphen{}1\_1}.
\bibitem[63]{handbook/references:id79}
\sphinxAtStartPar
J. van de Wetering. ZX\sphinxhyphen{}calculus for the working quantum computer scientist. 2020. \sphinxhref{https://arxiv.org/abs/2012.13966}{arXiv:2012.13966}.
\bibitem[64]{handbook/references:id80}
\sphinxAtStartPar
R. Duncan, A. Kissinger, S. Perdrix, and J. van de Wetering. Graph\sphinxhyphen{}theoretic Simplification of Quantum Circuits with the ZX\sphinxhyphen{}calculus. \sphinxstyleemphasis{Quantum}, 4:279, 2020. \sphinxhref{https://doi.org/10.22331/q-2020-06-04-279}{doi:10.22331/q\sphinxhyphen{}2020\sphinxhyphen{}06\sphinxhyphen{}04\sphinxhyphen{}279}.
\bibitem[65]{handbook/references:id77}
\sphinxAtStartPar
L. Berent, L. Burgholzer, and R. Wille. Towards a SAT encoding for quantum circuits: A journey from classical circuits to Clifford circuits and beyond. In \sphinxstyleemphasis{International Conference on Theory and Applications of Satisfiability Testing}. 2022. \sphinxhref{https://arxiv.org/abs/2203.00698}{arXiv:2203.00698}, \sphinxhref{https://doi.org/10.4230/LIPIcs.SAT.2022.18}{doi:10.4230/LIPIcs.SAT.2022.18}.
\bibitem[66]{handbook/references:id2}
\sphinxAtStartPar
T. Häner and D. S. Steiger. 0.5 petabyte simulation of a 45\sphinxhyphen{}Qubit quantum circuit. In \sphinxstyleemphasis{Int\textquotesingle{}l Conf. for High Performance Computing, Networking, Storage and Analysis}. 2017. \sphinxhref{https://doi.org/10.1145/3126908.3126947}{doi:10.1145/3126908.3126947}.
\bibitem[67]{handbook/references:id3}
\sphinxAtStartPar
J. Doi, H. Takahashi, R. Raymond, T. Imamichi, and H. Horii. Quantum computing simulator on a heterogenous HPC system. In \sphinxstyleemphasis{Int\textquotesingle{}l Conf. on Computing Frontiers}, 85\textendash{}93. 2019. \sphinxhref{https://doi.org/10.1145/3310273.3323053}{doi:10.1145/3310273.3323053}.
\bibitem[68]{handbook/references:id4}
\sphinxAtStartPar
T. Jones, A. Brown, I. Bush, and S. C. Benjamin. QuEST and high performance simulation of quantum computers. In \sphinxstyleemphasis{Scientific Reports}. 2018. \sphinxhref{https://doi.org/10.1038/s41598-019-47174-9}{doi:10.1038/s41598\sphinxhyphen{}019\sphinxhyphen{}47174\sphinxhyphen{}9}.
\bibitem[69]{handbook/references:id5}
\sphinxAtStartPar
G. G. Guerreschi, J. Hogaboam, F. Baruffa, and N. P. D. Sawaya. Intel Quantum Simulator: a cloud\sphinxhyphen{}ready high\sphinxhyphen{}performance simulator of quantum circuits. \sphinxstyleemphasis{Quantum Science and Technology}, 5(3):034007, 2020. \sphinxhref{https://doi.org/10.1088/2058-9565/ab8505}{doi:10.1088/2058\sphinxhyphen{}9565/ab8505}.
\bibitem[70]{handbook/references:id6}
\sphinxAtStartPar
X.\sphinxhyphen{}C. Wu, S. Di, E. M. Dasgupta, F. Cappello, H. Finkel, Y. Alexeev, and F. T. Chong. Full\sphinxhyphen{}state quantum circuit simulation by using data compression. In \sphinxstyleemphasis{Int\textquotesingle{}l Conf. for High Performance Computing, Networking, Storage and Analysis}. 2019. \sphinxhref{https://doi.org/10.1145/3295500.3356155}{doi:10.1145/3295500.3356155}.
\bibitem[71]{handbook/references:id70}
\sphinxAtStartPar
I. L. Markov and \textbackslash{}relax Yaoyun. Shi. Simulating quantum computation by contracting tensor networks. \sphinxstyleemphasis{SIAM Journal on Computing}, 38(3):963\textendash{}981, 2008. \sphinxhref{https://doi.org/10.1137/050644756}{doi:10.1137/050644756}.
\bibitem[72]{handbook/references:id71}
\sphinxAtStartPar
B. Villalonga, S. Boixo, B. Nelson, C. Henze, E. Rieffel, R. Biswas, and S. Mandrà. A flexible high\sphinxhyphen{}performance simulator for verifying and benchmarking quantum circuits implemented on real hardware. \sphinxstyleemphasis{npj Quantum Information}, 2019. \sphinxhref{https://doi.org/10.1038/s41534-019-0196-1}{doi:10.1038/s41534\sphinxhyphen{}019\sphinxhyphen{}0196\sphinxhyphen{}1}.
\bibitem[73]{handbook/references:id72}
\sphinxAtStartPar
J. Brennan, M. Allalen, D. Brayford, K. Hanley, L. Iapichino, L. J. O\textquotesingle{}Riordan, M. Doyle, and N. Moran. Tensor Network Circuit Simulation at Exascale. In \sphinxstyleemphasis{International Workshop on Quantum Computing Software (QCS)}, 20\textendash{}26. IEEE, November 2021. \sphinxhref{https://doi.org/10.1109/QCS54837.2021.00006}{doi:10.1109/QCS54837.2021.00006}.
\bibitem[74]{handbook/references:id73}
\sphinxAtStartPar
T. Vincent, L. J. O\textquotesingle{}Riordan, M. Andrenkov, J. Brown, N. Killoran, H. Qi, and I. Dhand. Jet: Fast quantum circuit simulations with parallel task\sphinxhyphen{}based tensor\sphinxhyphen{}network contraction. \sphinxstyleemphasis{Quantum}, 6:709, 2022. \sphinxhref{https://doi.org/10.22331/q-2022-05-09-709}{doi:10.22331/q\sphinxhyphen{}2022\sphinxhyphen{}05\sphinxhyphen{}09\sphinxhyphen{}709}.
\bibitem[75]{handbook/references:id9}
\sphinxAtStartPar
G. F. Viamontes, I. L. Markov, and J. P. Hayes. Improving gate\sphinxhyphen{}level simulation of quantum circuits. \sphinxstyleemphasis{Quantum Information Processing}, 2(5):347\textendash{}380, 2003. \sphinxhref{https://doi.org/10.1023/B:QINP.0000022725.70000.4a}{doi:10.1023/B:QINP.0000022725.70000.4a}.
\bibitem[76]{handbook/references:id23}
\sphinxAtStartPar
A. Botea, A. Kishimoto, and R. Marinescu. On the complexity of quantum circuit compilation. In \sphinxstyleemphasis{Int\textquotesingle{}l Symp. on Combinatorial Search}. 2018. \sphinxhref{https://doi.org/10.1609/socs.v9i1.18463}{doi:10.1609/socs.v9i1.18463}.
\bibitem[77]{handbook/references:id44}
\sphinxAtStartPar
R. Drechsler. \sphinxstyleemphasis{Advanced Formal Verification}. Springer, 2004. \sphinxhref{https://doi.org/10.5555/1024203}{doi:10.5555/1024203}.
\bibitem[78]{handbook/references:id45}
\sphinxAtStartPar
D. Janzing, P. Wocjan, and T. Beth. “Non\sphinxhyphen{}identity check” is QMA\sphinxhyphen{}complete. \sphinxstyleemphasis{International Journal of Quantum Information}, 03(03):463\textendash{}473, 2005. \sphinxhref{https://doi.org/10.1142/S0219749905001067}{doi:10.1142/S0219749905001067}.
\bibitem[79]{handbook/references:id32}
\sphinxAtStartPar
N. Quetschlich, L. Burgholzer, and R. Wille. MQT Bench: Benchmarking Software and Design Automation Tools for Quantum Computing. \sphinxstyleemphasis{Quantum}, 7:1062, 2023. \sphinxhref{https://doi.org/10.22331/q-2023-07-20-1062}{doi:10.22331/q\sphinxhyphen{}2023\sphinxhyphen{}07\sphinxhyphen{}20\sphinxhyphen{}1062}.
\end{sphinxthebibliography}

\renewcommand{\indexname}{Index}
\footnotesize\raggedright\printindex
\end{document}